\documentclass[journal]{IEEEtran}
\usepackage{graphicx}
\usepackage{epstopdf}
\usepackage{cite}
\usepackage{amsfonts}
\usepackage{amssymb}
\usepackage{amsmath}
\usepackage{mathrsfs}
\usepackage{bm}
\usepackage{amsmath}
\usepackage{mdwmath}
\usepackage{mdwtab}
\usepackage[ruled,vlined,linesnumbered]{algorithm2e} 
\usepackage{subfigure}
\usepackage{float}

\newtheorem{remark}{Remark}

\usepackage{makecell,multirow,diagbox}

\begin{document}

\title{AI-Assisted Low Information Latency Wireless Networking}
\author{Zhiyuan Jiang, Siyu Fu, Sheng Zhou, Zhisheng Niu,~\IEEEmembership{Fellow,~IEEE}, Shunqing Zhang, and\\ Shugong Xu,~\IEEEmembership{Fellow,~IEEE}
\thanks{Z. Jiang, S. Fu, S. Zhang and S. Xu are with Shanghai University, Shanghai, China. S. Zhou and Z. Niu are with Tsinghua University, Beijing, China. The corresponding author is Sheng Zhou.
}}
\maketitle

\begin{abstract}
The 5G Phase-2 and beyond wireless systems will focus more on vertical applications such as autonomous driving and industrial Internet-of-things, many of which are categorized as ultra-Reliable Low-Latency Communications (uRLLC). In this article, an alternative view on uRLLC is presented, that information latency, which measures the distortion of information resulted from time lag of its acquisition process, is more relevant than conventional communication latency of uRLLC in wireless networked control systems. An AI-assisted Situationally-aware Multi-Agent Reinforcement learning framework for wireless neTworks (SMART) is presented to address the information latency optimization challenge. Case studies of typical applications in Autonomous Driving (AD) are demonstrated, i.e., dense platooning and intersection management, which show that SMART can effectively optimize information latency, and more importantly, information latency-optimized systems outperform conventional uRLLC-oriented systems significantly in terms of AD performance such as traffic efficiency, thus pointing out a new research and system design paradigm.
\end{abstract}

\section{Introduction}
\label{sec_intro}
The focus of wireless communication systems has shifted from content communications, e.g., voice and video which are mainly human-based, to machine-based sensing/control information communications. For cellular systems, the 1G-4G systems have been mainly designed for delivering human-based contents; whereas 5G and beyond systems find most of their revenue in machine-based vertical applications such as Vehicle-to-Everything (V2X) communications and wireless networked factory automation. This trend has generated enormous technology revolutions in wireless communications. In fact, two of the three main targets of 5G are set for Machine-Type Communications (MTC)---massive MTC and mission-critical MTC which is also known as ultra-Reliable Low-Latency Communications (uRLLC) \cite{bennis18}. In essence, the unique technical challenges brought by MTC are due to the nature of machines: they can be in massive amount or density (corresp. massive MTC) and their perceptions and reactions are much more time-sensitive than human (corresp. uRLLC). 
In particular, some novel applications have very stringent communication latency requirements. For example, high-level (level 4-5) autonomous driving usually requires status messages delivered within less than $10$ ms to enable cooperative vehicle maneuver, dense platooning and etc; To replace wired connections (such as industrial Ethernet) in factory automation, the closed-loop communication latency, which consists of sensory data collection, data processing at the Programmable Logic Controller (PLC) and control information dissemination, should be less than $5$ ms, together with very low delay jitter (within micro-seconds); The Tactile Internet even requires less than $1$ ms latency to enable applications with immersive perceptions such as remote robotic surgery. 
\begin{figure*}[!t]
	\centering    
	{\includegraphics[width=.8\textwidth]{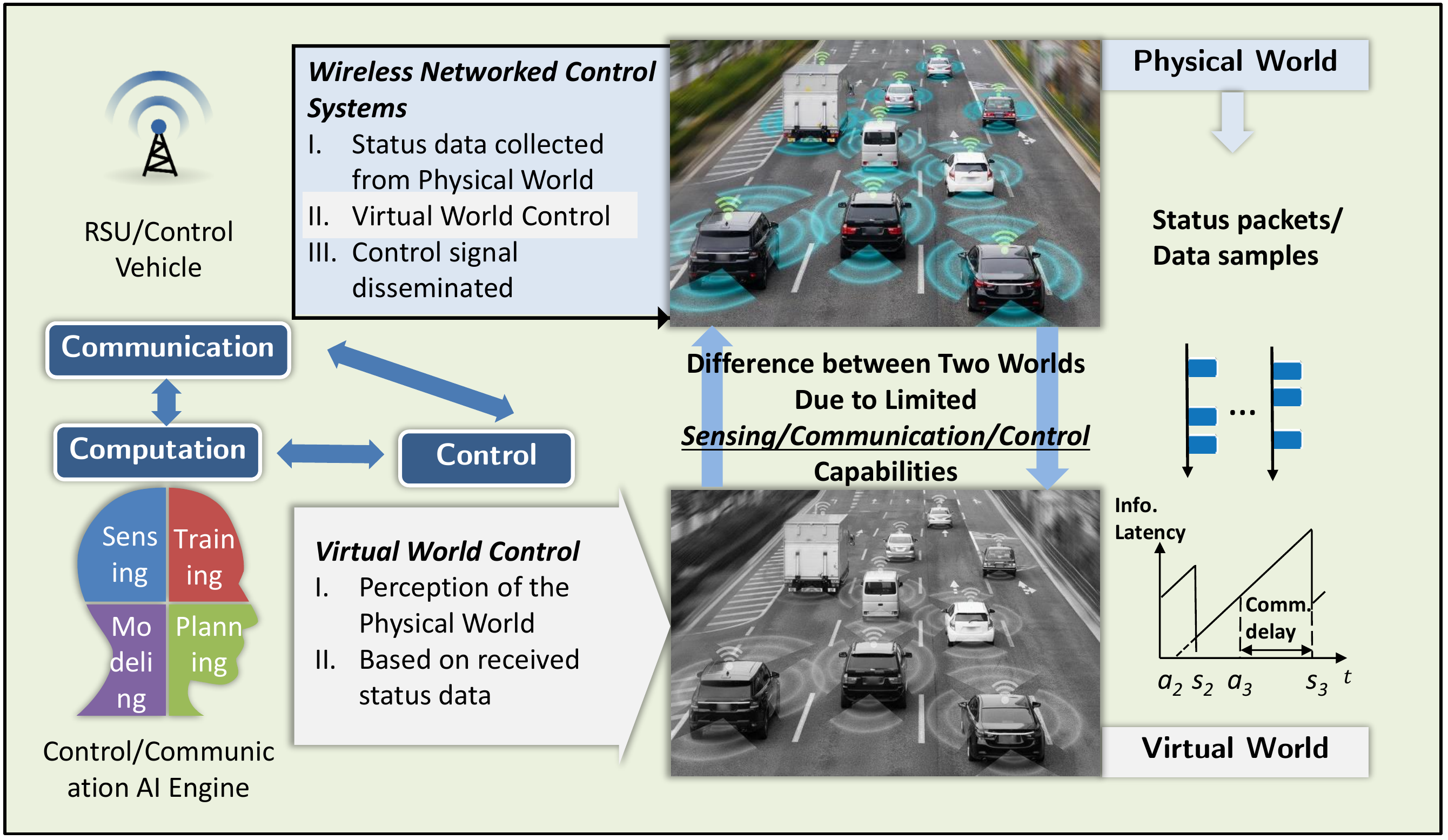}}
	\caption{A wireless control system employing virtual world abstractions of the physical world, wherein information latency plays a central role.}
	\label{fig_dt}
\end{figure*}

The currently standardized 5G system, somewhat surprisingly, has not yet tackled such uRLLC challenges. Despite the fact that some early evaluations report that 3GPP Release $15$ has achieved user plane latency of $1$ ms and control plane latency of less than $20$ ms, and that wireless protocols designed specifically for factory automation, such as Wireless Interface for Sensors and Actuators (WISA), can achieve a closed-loop latency less than $10$ ms, it should be noted that these results are obtained in highly idealized scenarios. For example, the system is lightly loaded for 5G tests, and dedicated time/frequency resources are allocated to terminals in WISA assuming no system dynamics. In practice, as widely recognized, there is no mystery in realizing uRLLC over air-interface, but to trade precious time/frequency/spatial resources for low-latency and reliability. Together with Mobile Edge Cloud (MEC) and network slicing technologies which lower application functionalities in the protocol stack to avoid core network delay, such an approach may provide uRLLC in certain scenarios, but sometimes with unaffordable resources. Therefore, the system \emph{scalability} to a large amount of terminals with dynamic traffic remains a severe issue. It is fair to say that great difficulties exist for wireless networks to support scalable, robust and more stringent uRLLC with limited resources in the future.

In this article, an alternative perspective for uRLLC to enable real-time control is proposed, that is an information latency perspective instead of communication latency. Based on this view, better application-based performance under limited wireless resources is achieved thanks to higher resource utilization efficiency. In what follows, this article first gives the definition of information latency metric and then shows that the metric is more directly related to control performance in real-world systems, thus optimizing it makes more sense for machine-based communications. The optimization of information latency essentially requires situational-awareness of terminals and hence calls for the aid of Artificial Intelligence (AI) to dynamically comprehend and recognize the situations (system states). Finally, we will provide concrete evaluation results of our framework, applied in high-level autonomous driving---one of the most intriguing and practical use cases for uRLLC.

\section{Wireless Networked Control Systems: Information Latency v.s. Communication Latency}
\label{sec_compare}
As shown in Fig. \ref{fig_dt}, many applications of MTC can be abstracted as Wireless Networked Control (WNC) systems, i.e., one or several control centers collect status information from distributed terminals through a wireless network, and then disseminates control signals, which are based on the collected status information and a specific control scheme, to actuators that would carry out the actions subsequently. Note that information flow and network topology may be diversified in practice, whereas this section focuses on a one-hop, single-control-unit scenario for ease of illustration. Typical applications include wireless networked factory automation wherein sensors and actuators are wirelessly controlled by a PLC; platooning in autonomous driving wherein the lead vehicle carries out the control decisions and sends to following vehicles. In WNC systems, control is based on perceived information about the physical world at the control unit, and the perceived information is obtained by wireless status update. The overall perceived status which the control relies on is usually referred to as the \emph{virtual world}, and it is clear that the control performance is highly related to the distortion between the virtual and physical worlds. The distortion stems from several aspects: limited sensing, communication and control capabilities, as well as imperfect modeling of the physical world. Limited sensing and communication capabilities are due to imperfect perception of the physical world by the devices, and communication latency (e.g., physical layer latency and scheduling latency) as well as packet loss, respectively. When the controller receives a data packet containing an update about the physical world, the status has already become stale, possibly with sensing errors. Imperfect control also results in distortion between the two worlds, for that the control effects of real-world objects are not always what we have expected or modeled, e.g., the real acceleration is different with the controlled value passed from the upper-controller to lower-controller of a vehicle due to gear shifting and air drag. 

Precisely modeling and analyzing the distortion is extremely challenging, whereas from a communication perspective, a feasible approach is to implicitly reflect the distortion by a latency metric that is intrinsically different with the conventional communication latency metric. Define the \textit{information latency} between the virtual and physical worlds as some rational measure (e.g., average or risk-sensitive measure such as the probability of exceeding a certain bound) of the time periods elapsed since the last true statuses of the physical world are observed and input to the virtual world accurately. Note that by this definition, the distortion intuitively increases with the information latency since the physical world is constantly changing. When the information latency is zero, that is, when the physical world is completely synchronized with the statuses of the virtual world, the virtual world equals the physical world.

\begin{table*}[!t]
	\centering
	\caption{Summary of State-of-the-art on AoI Optimization in Wireless Networks}
	\label{table_review}
	\begin{tabular}{|c|c|c|c|c|c|}
		\hline
		Ref. & Network Topology & MAC Protocol & Packet Arrivals & PHY Abstractions & Optimization Variables \\ 
		\hline
		\cite{kadota18} & \multirow{8}*{Star network} & \multirow{3}*{Scheduled access} & Active sources & i.i.d. error & \multirow{4}*{Scheduling}  \\
		\cline{1} \cline{4-5}
		\cite{hsu18} &  &  & Bernoulli & Ideal &   \\
		\cline{1} \cline{4-5}
		\cite{he17} &  &  & Buffered sources & Physical model &   \\
		\cline{1} \cline{3} \cline{4-5}
		\cite{jiang18_iot} &  & Round-robin & \multirow{2}*{Bernoulli } & \multirow{2}*{Ideal} &  \\
		\cline{1} \cline{3} \cline{6}
		\cite{jiang18_itc} &  & Prioritized CSMA &  &  &  $p_{\mathsf{tx},n}(t)$ \\
		\cline{1} \cline{3-6} 
		\cite{yates17} &  & ALOHA  & Active sources & i.i.d. error & \multirow{2}*{$p_{\mathsf{tx},n}$} \\
		\cline{1} \cline{3-5}
		\cite{kosta18} & & CSMA and round-robin & Bernoulli/Active sources   &  \multirow{3}*{Ideal}  &   \\
		\cline{1} \cline{3-4} \cline{6}
		\cite{ali19} &  & CSMA & Poisson &  & Backoff window  \\
		\cline{1-4} \cline{6}
		\cite{talak18} & \multirow{2}*{General network} & \multirow{2}*{ALOHA} & Active/Buffered sources & & $p_{\mathsf{tx},n}$  \\
		\cline{1} \cline{4-6}		
		\cite{talak_mobihoc} &  &  & Active sources & i.i.d. error & $p_{\mathsf{tx},n}$ and packet arrival rates  \\	
		\hline
	\end{tabular}
\end{table*}

Since information latency is directly related to the information timeliness observed at the control decision-maker, it is more reasonable for a wireless network to optimize information latency than communication latency if its end goal is optimizing the performance of a WNC system. Note that the concept of Age of Information (AoI) that has been gaining momentum in the literature can be included in the information latency definition. In particular, this definition is similar with the non-linear AoI \cite{kosta17} (or value of information) definition, with distinction that it is formulated under the physical-virtual world framework in this article. Unlike communication latency, information latency accounts for not only the end-to-end latency, but also communication reliability (unsuccessful updates cannot improve the information latency), sensing latency, information processing latency and all aspects of the information acquisition process that adds to the end effect of information latency. Therefore, it can summarize the interplay and impacts on control performance of several metrics that are intrinsically tradeoff between each other, e.g., communication delay and reliability given limited radio resources, or sensing latency/frequency and network queuing latency. This essentially makes information latency a flow-level Quality-of-Service (QoS) metric, and wireless network design in the MTC era is indeed suitable for such a metric, for that one distinctive feature of MTC is that packets transmitted by a specific radio over time are concerned with the same (or the same set of) information process. For example, a temperature sensor transmits packets about its sensed temperature over time; a vehicle in a platoon transmits its statuses, consisting of, e.g., speed, acceleration and spacing to the front vehicle, to the control vehicle. In view of this, a flow-level metric is reasonable to be considered as the QoS metric for MTC applications with such specified radio interfaces. 

\begin{remark}
	Note that the motivation to look at information latency relies on the assumption that supporting advanced, robust and scalable uRLLC with \emph{limited resources} is infeasible in some use cases. In other words, if the wireless network can provide such uRLLC for every packet, irrespective of the communicated status, terminal density and packet frequency, the information latency requirement would be automatically satisfied. However, as mentioned before, this condition is impractical in many scenarios, thus necessitating the research of information latency.
\end{remark}

%

\section{Information Latency Optimization in Wireless Networks: Overview and Challenges}
\label{sec_rw}
As discussed in the previous section, information latency is more related to the WNC performance than the conventional communication latency. Its optimization is increasingly important in future wireless systems. As we will show in Section \ref{sec_cs}, information latency optimization leads to better control performance in typical 5G vertical applications. 

Towards this end, several recent works have been dedicated to information latency optimizations in wireless networks, mostly considering AoI and focusing on Medium Access Control (MAC) layer operations---those have been summarized in Table \ref{table_review}. In a nutshell, it has been found out that the conventional MAC protocols, e.g., Carrier-Sense Multiple-Access (CSMA) and ALOHA, which are used in both IEEE 802.11 series and the random access procedure in LTE and 5G New Radio (NR), are unsuitable for AoI optimizations, at least when implemented in their original forms. Much of the work has focused on the most common star-topology network, wherein a set of terminals communicate with a common destination. Depending on whether the scheduling operations are centralized or decentralized, the researchers have proposed scheduled and uncoordinated access approaches which are suitable for the wireless downlink (corresp. centralized) and uplink (corresp. decentralized), respectively. Specifically, Ref. \cite{kadota18,hsu18,jiang18_itc} have adopted a Whittle's index approach which calculates a scalar value (i.e., index) representing the transmission urgency for each terminal based on its dynamic situations (e.g., AoI), and lets the most urgent terminal to transmit in each time slot. This index-based approach is shown to be near-optimal, supported by numerous evaluation results although not theoretically proven. For theoretical optimality, it is shown by \cite{jiang18_iot} that a round-robin scheduling scheme which requires small signaling overhead, is asymptotically optimal with a large number of terminals under certain conditions such as no transmission failures. In \cite{jiang18_itc,yates17,kosta18,ali19}, another important aspect regarding the uplink uncoordinated access design is investigated. Unlike scheduled access, terminals have to decide whether to transmit in each time slot autonomously and efficiently. Optimizations over the transmit probability (i.e., $p_{\mathsf{tx},n}$ where $n$ is the terminal index), or equivalently the contention backoff window size, in CSMA/ALOHA are carried out in \cite{yates17,kosta18,ali19}, considering terminals may have different channel conditions, service rates and packet arrival rates, respectively; however, the transmission probability is constant over time once decided in these works. In \cite{jiang18_itc}, dynamic access probability over time (i.e., $p_{\mathsf{tx},n}(t)$) according to the terminals' states is proposed, adopting the index approach to prioritize terminals. 

For more general network topology, Ref. \cite{talak18,talak_mobihoc} show a stationary decentralized policy, i.e., terminals transmit with constant probabilities over time. The transmit probability, that is optimized based on terminals' individual conditions, is optimal considering active sources and peak AoI. Furthermore, when considering the average AoI, it is within an order of $2$-optimal. However, it is notable that these order bounds for average AoI are generally quite loose. Here active sources mean that terminals transmit their instantaneous status whenever scheduled without considering the sensing latency; buffered sources are those store the arrival packets, which are useful when the whole history of status variation is of interest; in other cases, most works assume that the status packets arrive based on the Bernoulli (time-slotted systems) or Poisson (continuous-time systems) distribution and only keeps the most up-to-date packet in the queue. Physical-layer (PHY) abstractions also differ in these works, mostly assuming ideal channels or i.i.d. channel error.


\emph{Miscellaneous}:  Optimizing the information latency instead of AoI is considered as a step forward for status update. Ref. \cite{sun17_tit} investigates the optimal sampling policy to minimize the distortion with the communication part modeled as a memory-less service process. Whereas Ref. \cite{jiang19_infocom} jointly considers sensing and communication scheduling, and a mean-field based approach is hence proposed to address the decentralized status update issue.

A key distinguishing factor for information latency optimization in wireless networks, compared with communication latency, is that information latency is an end metric that usually relates to many dynamic aspects of the system, such as channel conditions, packet arrival patterns, status variations of terminals, network congestion conditions and so on. Many of these features are not well-modeled in the current literature, and often they have some time-domain structures, e.g., time correlation or Markovianity, that are hard to capture. These features are referred to as \emph{situations} hereinafter. The analysis and optimization of information latency are considered much more challenging than communication latency considering such dynamic, semantic situations. Specifically, despite the research progress mentioned above, three distinct challenges still exist: 
\begin{itemize}
	\item Complex, high-dimensional system state and action space in realistic scenarios;
	\item Self-optimized and robust design towards adaptivity to system dynamics and environment changes;
	\item Scalability considering meta information and signaling exchange overhead.
\end{itemize}

In the next section, we will elaborate on the key challenges and then exploit state-of-the-art Artificial Intelligence (AI) techniques to address these issues.

\section{AI-Assisted Situationally-Aware Wireless Networking}
\label{sec_main}
Modern AI techniques, especially deep learning, are tremendously successful mainly due to their powerful representation ability, which can be leveraged in our framework to tackle the challenge of complicated state and action space. On the other hand, the remaining two challenges for low information latency wireless networking are more unique that call for novel solutions. 
\subsection{Self Optimized Network: Adaptive without a Supervisor}
The concept of Self Optimized Network (SON) has been proposed for many years. However, it is not until recently with the emergence of AI that SON is promising to be truly realized. In the context of information latency optimization, the network has to be self-optimized towards situational-awareness, i.e., terminals in the network need to learn to comprehend the situation they are in given their own perceptions and make decisions about wireless transmissions accordingly, e.g., whether to transmit and how much time/frequency/spatial resources to use. Meanwhile, it is important to note that this learning procedure, unlike widely-adopted and greatly successful supervised learning, is necessary to be unsupervised or based on indirect supervision such as reinforcements. This is because obtaining an optimized dynamic solution to supervise the training of the network is difficult, since the situation and context information can be quite complex. On the other hand, online adaptivity and scalability to environment and network changes are mandatory because real-world scenarios are often time-varying and unpredictable---and similarly, these have to be achieved autonomously without a supervisor. 

\subsection{Distributed Intelligence with a Common Objective}
The other significant aspect for situational-awareness in the network concerns with the meta information (information facilitating transmit decisions) availability and hence scalability. In most scenarios, it is unreasonable to assume that terminals have access to all necessary meta information of the whole network. For example, individual and distributed terminals are unaware of the situations other terminals are in; moreover, exchanging this information among terminals entails prohibitive high signaling overhead. Therefore, distributed intelligence needs to be enabled. A key challenge here is that although the meta information is distributed and decentralized, terminals have to cooperate in some sense to achieve a common goal, whereas being selfish usually results in sub-optimal or even disastrous performance. One simple instance is wireless multiaccess networks wherein selfish terminals lead to congesting the channel all the time without any successful packet delivery. In this case, the distributed terminals should be properly incentivized to sacrifice their own interests, but to contribute to the common objective, e.g., optimizing the average utility over all terminals. This should be achieved by carefully designing a protocol based on which credit assignment is carried out by a central controller with minimal signaling overhead. And it brings about another major challenge of credit assignment---generally a big issue in multi-agent reinforcement learning---that is how to appropriately assign credits to terminals based on their individual actions, and how to evaluate these actions.

\subsection{SMART: Situationally-Aware Multi-Agent Reinforcement Learning Framework for Wireless Networks}

\begin{figure}[!t]
	\centering    
	{\includegraphics[width=0.4\textwidth]{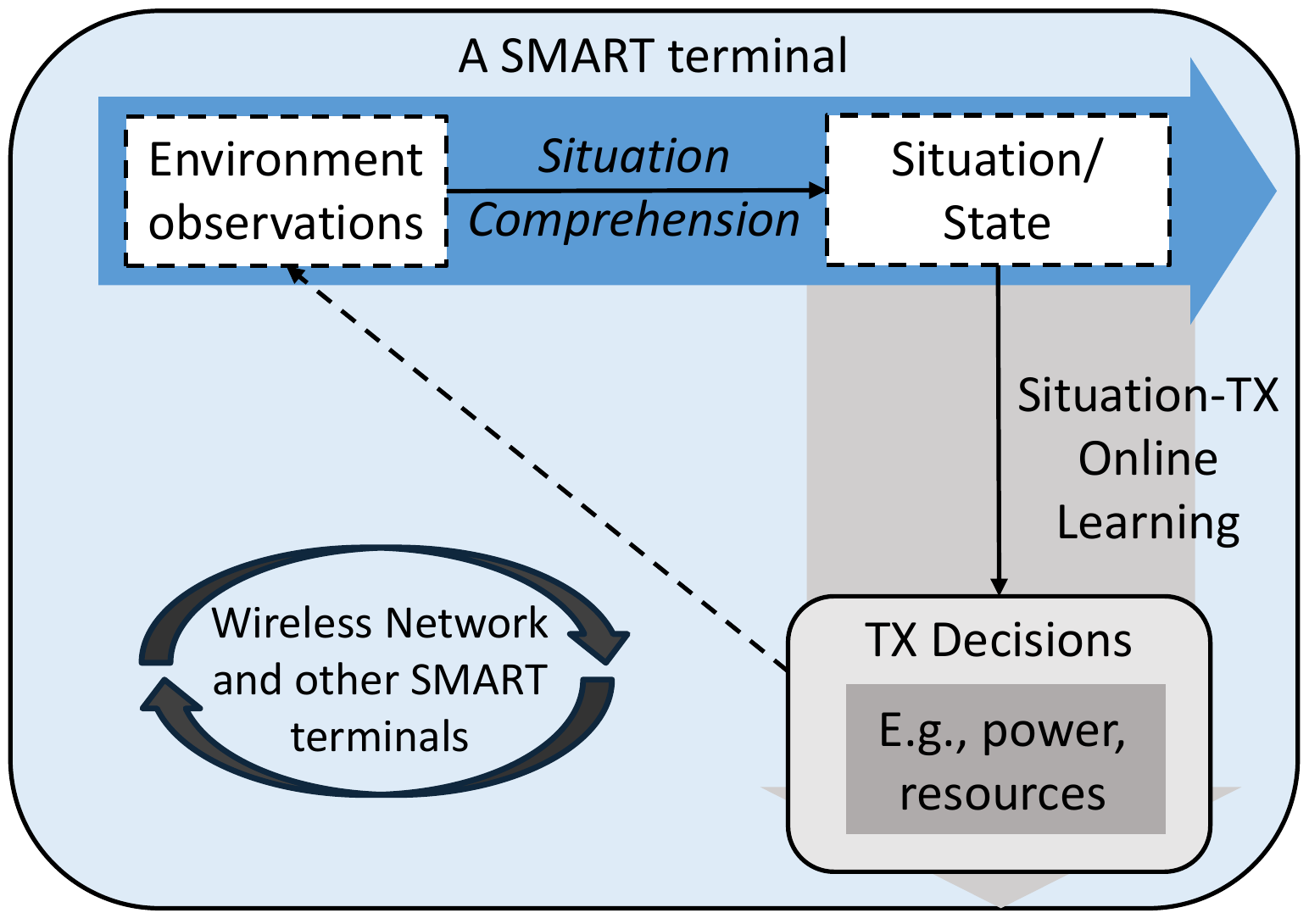}}
	\caption{High-level architecture of SMART, where the situation comprehension can be done offline to stabilize online learning process. The situation-transmission mapping learning should be online-adaptive.}
	\label{fig_smart}
\end{figure}

To realize self-optimized, adaptive and distributed situational-awareness for information latency optimization, we propose a SMART framework \cite{jiang19_spawc} based on multi-agent reinforcement learning, which is illustrated in Fig. \ref{fig_smart}. At each terminal in the network, a mapping function from the current situation of the terminal, which is modeled as the Markov state, to the transmit decision is learned. Such a decision could be about, e.g., transmit power, resources and so on, as long as it can be parameterized and hence represented. Similarly, the mapping function is also realized by parameterized approximations such as Deep Neural Networks (DNNs), whose tremendous success stems from strong representation ability and efficient, gradient-based back propagation training algorithms. 

\cite{jiang19_spawc}.
\begin{figure*}[!t]
	\centering    
	{\includegraphics[width=1\textwidth]{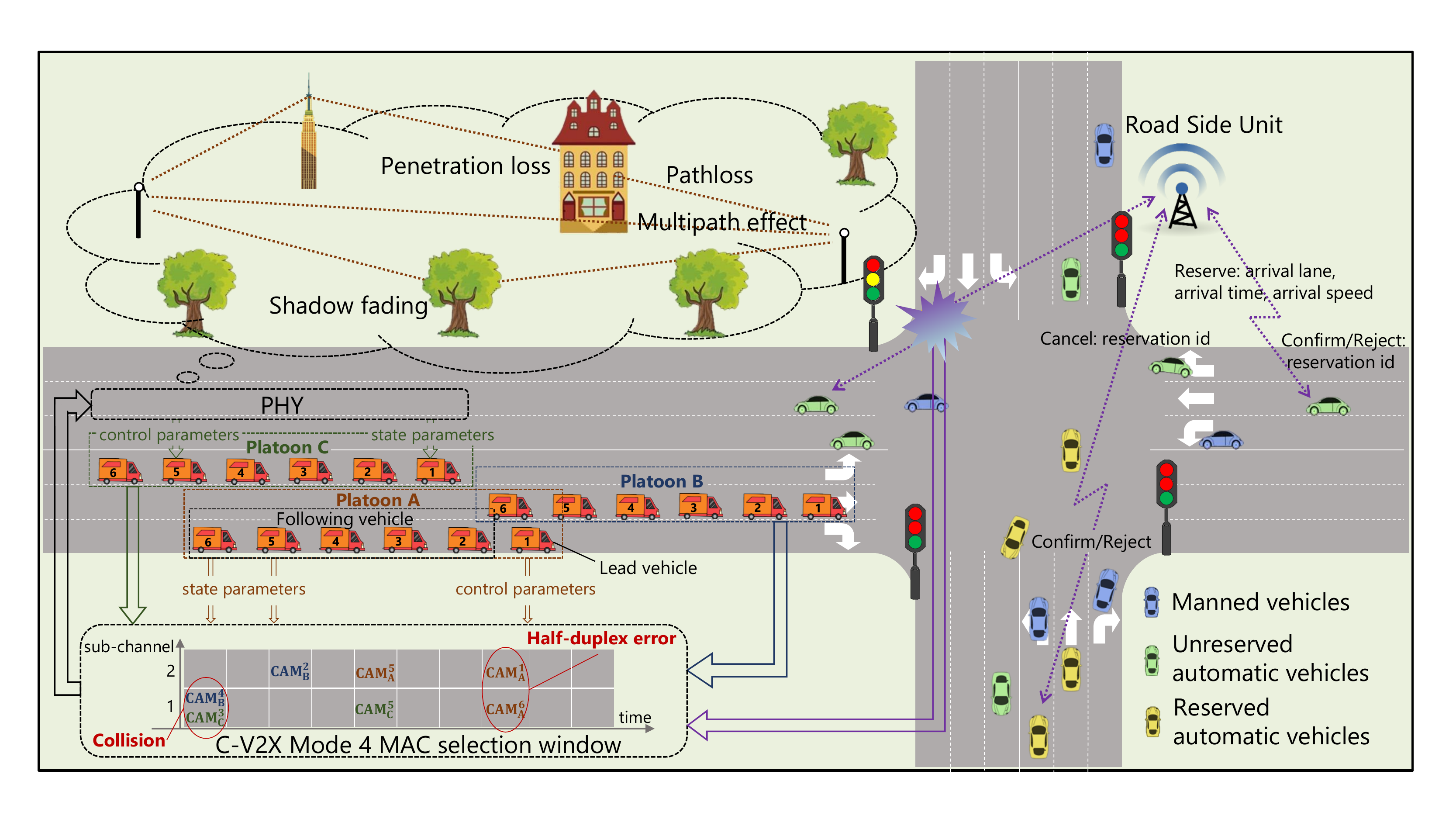}}
	\caption{Typical autonomous driving scenarios that are aided by V2X communications. Platooning and intersection management are shown in the figure, and messages are conveyed by LTE V2X Mode 4.}
	\label{fig_ad}
\end{figure*}
To address the issue of no supervisor, fortunately, the reinforcement learning framework can be leveraged. Thereby, instead of training the mapping function by a genie-aided and thus impractical optimal actions, a reward is fed to the learning algorithm which reflects how good the set of actions is to date---note that the reward may be delayed and hence the history of actions affects the current reward. Then comes the question: who should assign this reward over time so as to incentivize terminals towards achieving the common goal? In some wireless networks such as a wireless multiaccess network wherein terminals report to the same node, this problem is trivial since the central node can take this responsibility and the rewards (possibly different among terminals) can be broadcasted to all terminals. However, the selection of such a semi-supervisor is non-trivial in an ad-hoc network, by noting that feedback storm can easily occur if too many are selected to feedback the reward. 

Distributed situational-awareness requires terminals be able to intelligently make transmit decisions based on their local observations, which only makes sense when local observations can reflect the transmission urgency. This is true for information latency optimization since local status changes, channel variations are observable that affect the transmit decision. In contrast, the conventional communication latency optimization can be regarded as state-less, since packets are treated equally---one bit is one bit.
The key challenges for distributed situational-awareness with a common goal can be summarized as scalability and convergence. It is well known that multi-agent reinforcement learning algorithms do not scale well with the number of terminals, and may converge to a possibly bad solution. The root reason is that the observations of terminals are incomplete, and thus unlike full-information-state reinforcement learning which usually converges, partial information of terminals lead to learning towards a moving target. This is particularly troublesome when the number of terminals is large, wherein it is increasingly difficult for terminals which learn their own strategies regarding others as static environment to converge to an optimized solution. To address this issue, the SMART framework adopts a two-stage training process: a pre-training phase during which terminals in the network are pre-trained to comprehend their situations that are irrelevant of online network variations, and an online training phase that adapts to specific network conditions, topology and traffic. The design of such a decoupling approach has theoretical support in the case of wireless multiaccess networks, where the Whittle's index approach indicates that pre-training the terminals given an auxiliary, scalar subsidy that adapts to the network conditions results in near-optimal performance 

\section{Case Studies}
\label{sec_cs}
In this section, building on our existing works \cite{jiang18_iot,jiang19_spawc,jiang19_infocom}, we present applications of the proposed SMART framework in high-level autonomous driving---one of the major use cases of 5G uRLLC. Two scenarios are demonstrated, i.e., dense platooning and intersection management as shown in Fig. \ref{fig_ad}, which both require uRLLC to coordinate among vehicles efficiently. Meanwhile, both cases can be considered as real-time WNC systems, whose performance critically depends on information latency. 


We implement the dense platooning and intersection management control schemes on Simulation of Urban MObility (SUMO), as well as the wireless communication interface following the LTE V2X Release $14$ Mode 4 standard. Mode 4 adopts a semi-persistent scheduling scheme, based on which vehicles autonomously choose the transmission time/frequency resources. The carrier frequency is set to $5.9$ GHz and the bandwidth is $20$ MHz. The Winner B1 channel model is adopted, with the shadow fading coefficient following the log-normal distribution with a standard variance of $6$ dB. The transmit power is $23$ dBm. The modulation and coding scheme is QPSK and turbo coding with $\frac{1}{3}$ coding rate. The statuses are conveyed by messages with packet size of $300$ bytes, each occupies $1$ ms and $50$ frequency resource blocks ($10$ MHz, corresponding to $2$ sub-channels). A collision-based interference model is adopted that two or more packets occupying the same time and frequency sub-channel are assumed to incur a transmission error. 

For ease of exposition, in this section, we fix the air-interface latency to be $5$ ms on average (including distributed resource selection in Mode 4 and PHY latency), and study the tradeoff between the specific AD performance and transmission reliability. Note that in Mode 4, similar with many other wireless communication schemes, the reliability and communication latency are a tradeoff, and hence the results obtained by studying the impact of reliability given fixed latency can be transferred to understand both.
\subsection{Dense Platooning}
\begin{figure}[H]
	\centering    
	{\includegraphics[width=0.47\textwidth]{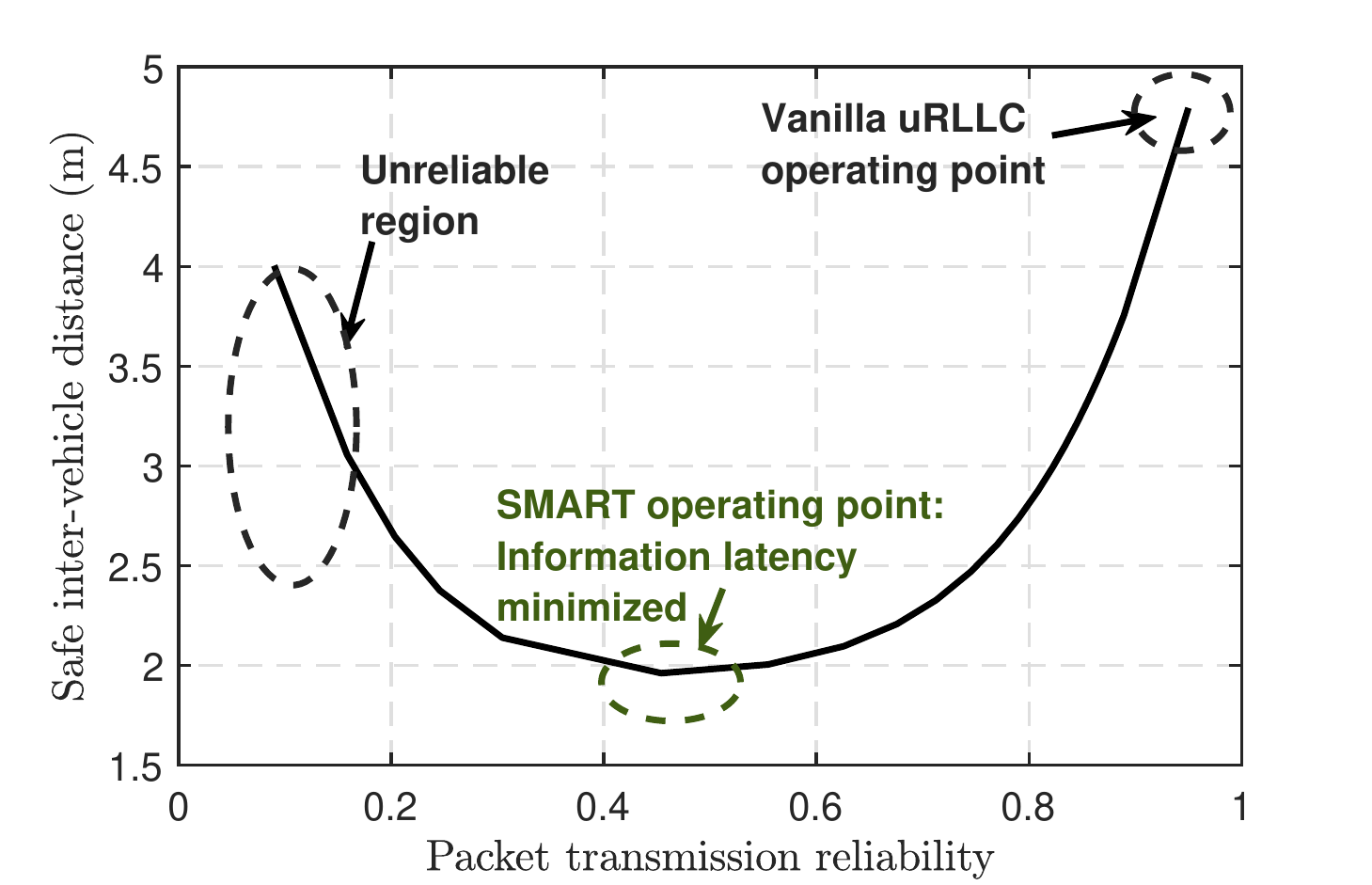}}
	\caption{Safe inter-vehicle distance tested on SUMO. Each platoon is controlled by its lead vehicle, and the status and control signaling are communicated based on LTE V2X Mode 4.}
	\label{fig_platoon}
\end{figure}

Platooning is a high-level autonomous driving technology that allows several vehicles to form a platoon, such that vehicles except the lead do not need a human driver. One distinct merit is fuel reduction, since the air drag of following vehicles is reduced, which depends on the fact that dense platooning is enabled when the inter-vehicle distance is reduced to less than, e.g., $5$ meters. This becomes very challenging with high-mobility, when communication-free solutions become unstable because they need constant head-time for collision avoidance. Therefore, the vehicle-to-vehicle communications are crucial to enable constant head-distance dense platooning, such that the air drag is further reduced while guaranteeing safety.
We simulate $8$ lanes on SUMO, and $6$ platoons which each consists of $8$ vehicles. The lead vehicle of each platoon first travels at a constant speed of $22$ m/s, then brakes with deceleration of $-2.94$ m/s$^2$, and then accelerates at $2$ m/s$^2$. During this process, the following vehicles in each platoon are controlled by the lead vehicle while reporting their statuses (speed, distance to front vehicle and acceleration), both through the PC5 interface (i.e., LTE V2X Mode 4). We consider a sensing refresh interval of $60$ ms and an actuation delay of $10$ ms which denotes the actuation time lag of the lower controller, i.e., the time between receiving the control signal and actually carrying out the action. The simulation is ran for $100$ times on SUMO. The minimum safe distance is defined as the minimum inter-vehicle distance that all vehicles do not crush during the $100$ runs. SMART is implemented based on minimizing the information latency by adjusting the message update rate in a distributed manner. Fig. \ref{fig_platoon} shows that, after convergence, SMART operates at the point where the packet transmission success rate is about $50\%$, resulting in the minimum safe inter-vehicle distance. However, vanilla uRLLC requires ultra-high packet transmission reliability, which is realized by, for lacking of better choices, many repetitive packets. This paradigm is shown to be inefficient, since repeating the same stale packet is clearly sub-optimal---in fact, the resultant performance is disastrous.
\begin{figure}[!t]
	\centering    
	{\includegraphics[width=0.47\textwidth]{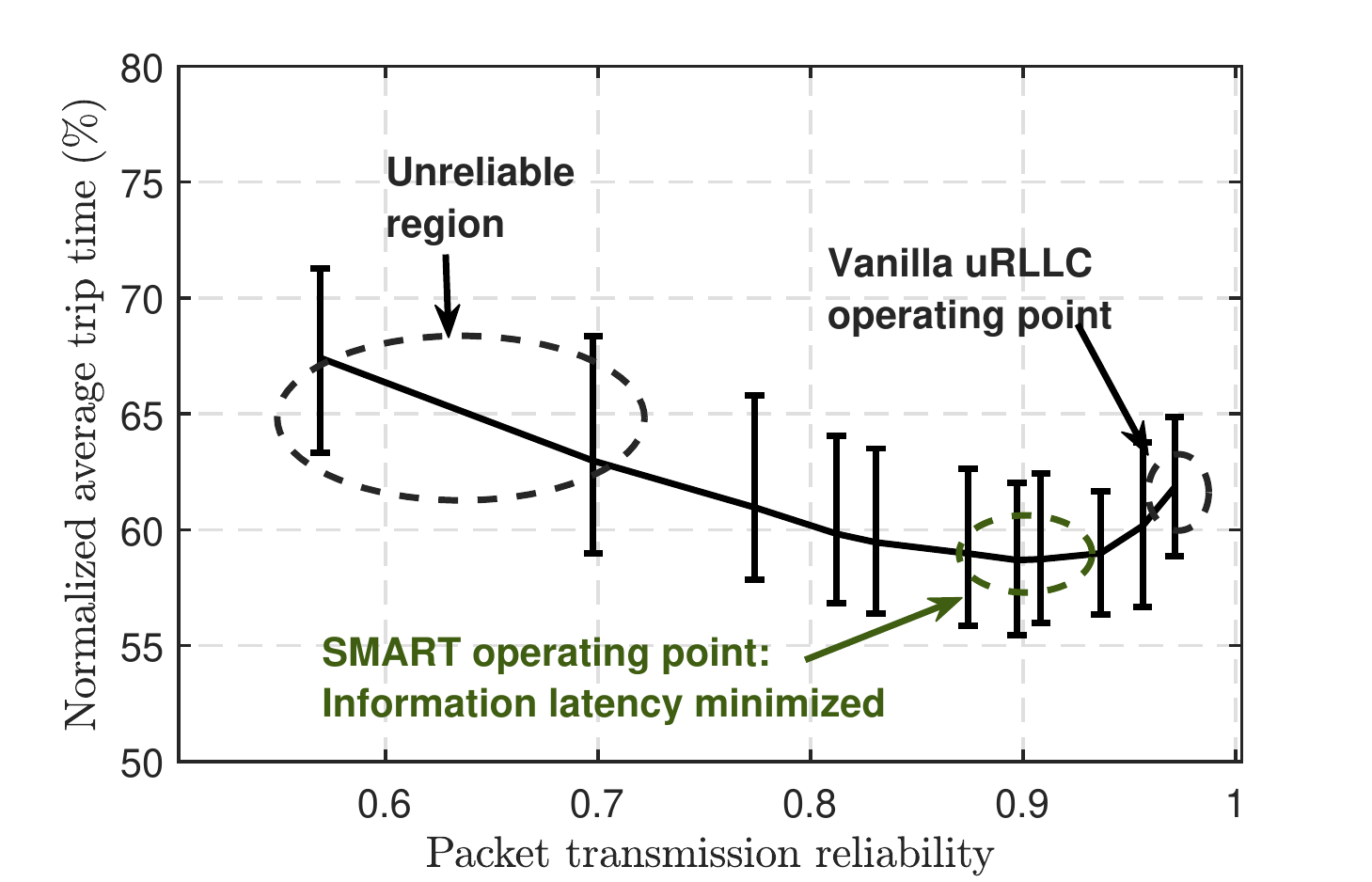}}
	\caption{Normalized average trip time tested on SUMO. Vehicles which are passing through an intersection reserve intersection resources from an RSU controller, and the status and control signaling are communicated based on LTE V2X Mode 4.}
	\label{fig_fcfs}
\end{figure}

\subsection{Intersection Management}
Another important AD scenario which V2X communications can benefit significantly is the intersection. Individual vehicle intelligence based on sensing is insufficient to handle the complicated traffic conditions and high requirements of coordination among vehicles. We simulate a hybrid AD scenario, wherein Human-driver Vehicles (HVs, $10\%$ in the simulations) and fully-Autonomous Vehicles (AVs, $90\%$) coexist. A four-way intersection ($40$ m $\times$ $40$ m) on SUMO is selected. HVs still follow the traffic lights, whereas AVs are controlled by the Road Side Unit (RSU) without following the traffic lights, but always give priorities to HVs. All vehicles report their locations, speeds and intentions to the RSU periodically. AVs go through the intersection based on time-spatial resources reservations coordinated by the RSU, i.e., AVs send request of a certain trajectory and status report (speed to intersection and lane information) to the RSU before going into the intersection and the RSU coordinates among vehicles and then feeds back the confirm or reject signals. 
We simulate $520$ vehicles passing through in each run and $100$ runs. The normalized average trip time of a vehicle is shown in Fig. \ref{fig_fcfs}. The $y$-axis is normalized by the average trip time controlled based on a traditional traffic light system. It is observed that AD can definitely improve the traffic efficiency. Similar with platooning, conventional uRLLC cannot provide optimal performance while SMART achieves higher traffic efficiency, although the improvement is smaller compared with platooning. This is because information latency introduced by wireless communications (usually in the range of tens of milliseconds) is more critical for control-level AD, e.g., platooning wherein vehicles accelerations are controlled by the communicated signaling, than planning-level AD, e.g., commonly-used intersection management where vehicles make reservations about the time-spatial resources before entering the intersection. 

A common observation from both case studies is that when updating status information, it is no longer important about individual packet delay and reliability, but critical for information update latency to be minimized. In some sense, it is better to be timely and unreliable---in which case any time the transmission is successful, the packet contains timely information---than to be ultra-reliable but compromising the information freshness.

\section{Conclusions and Outlook}
\label{sec_cl}
This article demonstrates that information latency is more critical than conventional uRLLC metrics in typical 5G vertical applications. Towards this end, SMART is proposed for information latency optimizations in wireless networks with 1) Self-optimization and online adaptivity; 2) Scalability with decentralized implementation. SMART is tested for typical applications in high-level AD such as dense platooning and intersection management. The results show that SMART can effectively minimize information latency, and more importantly, information latency optimized control is superior to conventional uRLLC-oriented systems in terms of the ultimate application performance, inspiring us to rethink the uRLLC design principle.

Many intriguing research directions exist. The scalability and convergence issue remains a key challenge for distributed learning in wireless systems, along with the credit assignment scheme to accelerate learning in a multi-agent setting. The semi-supervisor election problem in an ad hoc network is also relevant considering stability. Last but not least, a novel network architecture that supports AI-assisted, situationally-aware wireless systems is yet to come.

\bibliographystyle{ieeetr}
\bibliography{sm}

\begin{thebibliography}{10}

\bibitem{bennis18}
M.~{Bennis}, M.~{Debbah}, and H.~V. {Poor}, ``Ultrareliable and low-latency
  wireless communication: Tail, risk, and scale,'' {\em Proceedings of the
  IEEE}, vol.~106, pp.~1834--1853, Oct 2018.

\bibitem{kadota18}
I.~Kadota, A.~Sinha, E.~Uysal-Biyikoglu, R.~Singh, and E.~Modiano, ``Scheduling
  policies for minimizing age of information in broadcast wireless networks,''
  {\em IEEE/ACM Trans. Netw.}, vol.~26, pp.~2637--2650, Dec. 2018.

\bibitem{hsu18}
Y.~{Hsu}, ``Age of information: Whittle index for scheduling stochastic
  arrivals,'' in {\em IEEE Int'l Symp. Info. Theory}, pp.~2634--2638, Jun.
  2018.

\bibitem{he17}
Q.~{He}, D.~{Yuan}, and A.~{Ephremides}, ``Optimal link scheduling for age
  minimization in wireless systems,'' {\em IEEE Trans. Inform. Theory},
  vol.~64, pp.~5381--5394, Jul. 2018.

\bibitem{jiang18_iot}
Z.~{Jiang}, B.~{Krishnamachari}, X.~{Zheng}, S.~{Zhou}, and Z.~{Niu}, ``Timely
  status update in wireless uplinks: Analytical solutions with asymptotic
  optimality,'' {\em IEEE Internet of Things Journal}, vol.~6, pp.~3885--3898,
  Apr 2019.

\bibitem{jiang18_itc}
Z.~{Jiang}, B.~{Krishnamachari}, S.~{Zhou}, and Z.~{Niu}, ``Can decentralized
  status update achieve universally near-optimal age-of-information in wireless
  multiaccess channels?,'' in {\em International Teletraffic Congress (ITC
  30)}, vol.~01, pp.~144--152, Sep. 2018.

\bibitem{yates17}
R.~D. Yates and S.~K. Kaul, ``Status updates over unreliable multiaccess
  channels,'' in {\em IEEE Int'l Symp. Info. Theory}, pp.~331--335, Jun 2017.

\bibitem{kosta18}
A.~Kosta, N.~Pappas, A.~Ephremides, and V.~Angelakis, ``Age of information
  performance of multiaccess strategies with packet management,'' {\em arXiv
  preprint arXiv:1812.09201}, 2018.

\bibitem{ali19}
A.~Maatouk, M.~Assaad, and A.~Ephremides, ``Minimizing the age of information
  in a {CSMA} environment,'' {\em arXiv preprint arXiv:1901.00481}, 2019.

\bibitem{talak18}
R.~{Talak}, S.~{Karaman}, and E.~{Modiano}, ``Distributed scheduling algorithms
  for optimizing information freshness in wireless networks,'' in {\em IEEE
  Int. Workshop Signal Process. Adv. Wireless Commun. (SPAWC)}, pp.~1--5, Jun
  2018.

\bibitem{talak_mobihoc}
R.~Talak, S.~Karaman, and E.~Modiano, ``Optimizing information freshness in
  wireless networks under general interference constraints,'' in {\em ACM Int.
  Symp. Mobile Ad Hoc Netw. Comput. (MobiHoc)}, pp.~61--70, 2018.

\bibitem{kosta17}
A.~Kosta, N.~Pappas, A.~Ephremides, and V.~Angelakis, ``Age and value of
  information: Non-linear age case,'' in {\em IEEE Int'l Symp. Info. Theory},
  pp.~326--330, Jun 2017.

\bibitem{sun17_tit}
Y.~{Sun}, E.~{Uysal-Biyikoglu}, R.~D. {Yates}, C.~E. {Koksal}, and N.~B.
  {Shroff}, ``Update or wait: How to keep your data fresh,'' {\em IEEE Trans.
  Inform. Theory}, vol.~63, pp.~7492--7508, Nov 2017.

\bibitem{jiang19_infocom}
Z.~Jiang, S.~Zhou, Z.~Niu, and Y.~Cheng, ``A unified sampling and scheduling
  approach for status update in wireless multiaccess networks,'' in {\em IEEE
  Conf. Comput.Commun. (INFOCOM)}, May 2019.

\bibitem{jiang19_spawc}
Z.~Jiang, A.~Marinescu, L.~DaSilva, S.~Zhou, and Z.~Niu, ``Scalable multi-agent
  learning for situationally-aware multiple-access and grant-free
  transmissions,'' in {\em IEEE Int. Workshop Signal Process. Adv. Wireless
  Commun. (SPAWC)}, June 2019.

\end{thebibliography}
\end{document}